\documentclass[conference]{IEEEtran}
\IEEEoverridecommandlockouts
\usepackage{cite}
\usepackage{amsmath,amssymb,amsfonts}
\usepackage{algorithmic}
\usepackage{graphicx}
\usepackage{graphics}
\usepackage{textcomp}
\usepackage{xcolor}
\usepackage{subcaption}
\usepackage{pbox}
\usepackage{cleveref}
\usepackage{float}
\usepackage{inconsolata}
\usepackage{listings}
\definecolor{javapurple}{rgb}{0.5,0,0.35}
\definecolor{javagreen}{rgb}{0,0.4,0}
\lstdefinelanguage{customc}
{
  morekeywords={for, int, return},
  morecomment=[l]{//},
  morecomment=[s]{/*}{*/},
  morestring=[b]",
  basicstyle=\small\ttfamily,
  numbers=left, firstnumber=1, numberstyle=\tiny\color{gray},
  showstringspaces=false,
  escapeinside={(*@}{@*)},
  keywordstyle=\color{javapurple},
  columns=fullflexible,
  showlines=true,
  xleftmargin=0.5cm,
  moredelim=**[is][\color{NavyBlue}]{!}{!},
  moredelim=**[is][\color{OliveGreen}]{<_}{_>}
}
\lstdefinelanguage{python}{
  morekeywords={if, else, def, return},
  morecomment=[l]{//},
  morecomment=[s]{/*}{*/},
  morestring=[b]",
  basicstyle=\small\ttfamily,
  numbers=left, firstnumber=1, numberstyle=\tiny\color{gray},
  showstringspaces=false,
  escapeinside={(*@}{@*)},
  keywordstyle=\color{javapurple},
  columns=fullflexible,
  showlines=true,
  xleftmargin=0.5cm,
  moredelim=**[is][\color{NavyBlue}]{!}{!},
  moredelim=**[is][\color{OliveGreen}]{<_}{_>}
}

\def\BibTeX{{\rm B\kern-.05em{\sc i\kern-.025em b}\kern-.08em
    T\kern-.1667em\lower.7ex\hbox{E}\kern-.125emX}}
\begin{document}

\title{Code Transpilation for Hardware Accelerators}
\author{\IEEEauthorblockN{Yuto Nishida, Sahil Bhatia, Shadaj Laddad, Hasan Genc, Yakun Sophia Shao, Alvin Cheung}
\IEEEauthorblockA{\textit{UC Berkeley} \\
\{sahilbhatia, shadaj\}@berkeley.edu}
\vspace{-0.25in}
}

\maketitle

\begin{abstract}
DSLs and hardware accelerators have proven to be very effective in optimizing computationally expensive workloads. In this paper, we propose a solution to the challenge of manually rewriting legacy or unoptimized code in domain-specific languages and hardware accelerators. We introduce an approach that integrates two open-source tools: \emph{Metalift}, a code translation framework, and \emph{Gemmini}, a DNN accelerator generator. The integration of these two tools offers significant benefits, including simplified workflows for developers to run legacy code on Gemmini generated accelerators and a streamlined programming stack for Gemmini that reduces the effort required to add new instructions. This paper provides details on this integration and its potential to simplify and optimize computationally expensive workloads.

\end{abstract}


\section{Introduction}

In recent years, the software industry has witnessed a trend where Domain-Specific Languages (DSLs) have increasingly become part of the existing workflows. These specialized programming languages offer high-level abstractions that are tailored to solving particular problems or expressing computations within specific domains. Some examples of DSLs are Numpy for matrix operations, Tensorflow for deep learning, and Halide for image processing, among others. On the other hand, in the hardware industry accelerators have become increasingly popular. These special-purpose execution engines are optimized to perform specific tasks, and by offloading certain parts of computations to them while performing the remainder on a general-purpose CPU, applications can achieve performance optimizations.

One common outcome of these trends in both industries is the development of frameworks that facilitate the adoption of these specialized tools. By offering high-level abstractions and automating many low-level implementation details, these frameworks have made it easier for users to take advantage of the benefits of DSLs and accelerators.
One example of such a framework is \textbf{Metalift} \cite{metalift}, which enables users to build custom compilers for translating code written in general-purpose languages to DSLs. Leveraging program synthesis, Metalift frees developers from writing syntax-driven rules that can be error-prone and difficult to specify. On the hardware side, \textbf{Gemmini} \cite{gemmini} is an open-source framework for building custom DNN accelerators. It allows developers to generate accelerators and customize them end-to-end, from architectural templates (such as spatial arrays and scratchpads) to programming support (including ONNX format and low-level C++ APIs) to system support (such as microcontrollers and server-class CPUs).

At present, Gemmini offers two front-ends for running DNN workloads: high-level push-button support for executing workloads from ONNX files and low-level C/C++ APIs of its instruction set for running workloads on the generated accelerators. However, running legacy or unoptimized code, such as Fortran-based scientific computing kernels or C++-based image processing kernels on Gemmini could pose a challenge for developers. To do so, the developer would either need to manually translate these kernels or write a syntax-driven compiler to perform the translation, both of which can be error-prone and time-consuming. Combining Metalift with Gemmini can be a powerful approach for automating this translation from general-purpose languages to Gemmini's ISA. 
Additionally, Metalift's search-based technique for translation can be guided by the performance of the translated code on the generated hardware. This allows for the search to potentially find hardware parameters that are optimal for running a particular kernel.

This paper presents our initial work on combining Metalift and Gemmini. By integrating these two frameworks, we aim to enable more efficient computations in various domains and significantly simplify the workflow for developers. Additionally, since both frameworks are open-source it allows for greater flexibility in integrating the two frameworks which could lead to improvements and optimizations that benefit both the communities.
\section{Approach}
We apply Metalift to map array processing code written in standard Python or C++ (using standard lists and loops, rather than specialized library functions) to operations that can be accelerated using Gemmini. Instead of having to hand-write pattern matching logic to translate common patterns to their accelerated equivalents, which requires significant engineering and yields brittle results, we can focus on specifying the formal semantics of the Gemmini operators and let Metalift search for an appropriate mapping.

\begin{figure*}[t]
\centering
\begin{subfigure}{0.45\linewidth}
\begin{lstlisting}[language=customc,basicstyle=\ttfamily\small]
vector<int> program(vector<int> data){
    vector<int> result;
    for (int i = 0; i < data.size() - 1; i++)
        result.push_back(data[i] + data[i + 1]);
    return result;
}
\end{lstlisting}
\caption{User-Provided Sequential C++ Code}
\label{fig:c_code}
\end{subfigure}
\begin{subfigure}{0.5\linewidth}
\centering
\resizebox{\columnwidth}{!}{
\begin{tabular}{c c}
Initial Condition & \pbox{40cm}{\vspace{0.3cm}{$Inv(i=0, result=\{\}, data)$}\vspace{0.3cm}} \\
\hline
Preservation & \pbox{40cm}{\vspace{0.3cm} $Inv(i, \; result, \; data) \land (i \; < size(data)) \rightarrow $ \\ \vspace{0.3cm} $Inv( i \; + \; 1 \;, result.push\_back(data[i] + data[i+1]) \;, data)$} \\
\hline
Termination &  \pbox{40cm}{\vspace{0.2cm}$Inv(i, \; result, \; data) \land \neg \; (i \; < size(data)) \rightarrow PS(result \;, data)$}
\end{tabular}
}
\caption{Verification conditions for the source code.}
\label{fig:vc_proof}
\end{subfigure}

\hfill

\begin{subfigure}{0.45\linewidth}
\begin{lstlisting}[language=python,basicstyle=\ttfamily\small]
def gemmini_conv(data, kernel, stride):
    if length(data) < len(kernel):
        return []
    
    return prepend(
        dot_product(data, kernel), 
        gemmini_conv(data[stride:], kernel, stride))
\end{lstlisting}
\caption{Semantics of the convolution operator provided to Metalift}
\label{fig:conv}
\end{subfigure}
\begin{subfigure}{0.45\linewidth}
\begin{lstlisting}[language=python,basicstyle=\ttfamily\small,mathescape=true]
Synthesized: 
program = gemmini_conv(data, kernel=[1,1], stride=1)

Invariant: 
(i $\ge$ 0) && (i $<$ (length(data)) && 
     out = gemmini_conv(data[:i+1],kernel=[1,1],stide=1)
\end{lstlisting}
\caption{Program lifted to use Gemmini operators}
\label{fig:synth_res}
\end{subfigure}
\caption{The four steps of the Metalift pipeline, adapted for the domain of translating C++ kernels to Gemmini.}
\end{figure*}



The verified lifting framework that Metalift enables involves three key steps: defining a grammar to search for the target language, specifying the semantics of the target language, and analyzing the behavior of the source program we want to match the behavior of. The specifications provided to Metalift are written using a custom intermediate representation that resembles the APIs of SMT solvers such as Z3. 
To analyze the source program, Metalift provides several front-ends that perform static analysis over Python or LLVM inputs and generates a symbolic expression over the inputs in the same Metalift IR.

Metalift performs the software synthesis procedure with an iterative algorithm which enumerates candidate programs from the provided grammar, and then verifies the correctness of the candidate by querying an SMT solver. To perform this verification step for code involving loops with dynamic bounds, Metalift synthesizes additional \emph{loop invariants}, which make it possible to reason about the result of the program for any number of iterations.

Consider the input code in Figure~\ref{fig:c_code}, which computes the sums of values within a sliding window. Readers familiar with primitives for tensor accelerators may recognize this as a convolution, but this is not explicit in the code and the developer may not have this same insight. Our goal is to use verified lifting to automatically synthesize an provably equivalent function that uses an accelerated convolution operator.

Our first step using Metalift is to specify the grammar to search for the target program. We model each accelerated operator as a function that takes input tensors and configuration parameters as inputs and produces a new tensor as an output. Then, we can build up the grammar by starting with the inputs to the programs as leaves and layering compositions of the available Gemmini operators.

The other half of specifying the target language is to provide formal models of each operator, so that we can verify synthesized candidates against the source program using an SMT solver. In Metalift, the target language is specified by providing implementations in terms of the Metalift IR, which is later passed to the solver. In our implementation, we include specifications for core Gemmini operators such as matrix multiplication and convolutions (Figure~\ref{fig:conv}). But beyond these core specifications, Metalift is given no additional information about when these operators should be used---it has to discover appropriate mappings by searching the grammar.

\section{Results and Future Work}
Metalift can translate the code in \cref{fig:c_code} to generate the equivalent code in Gemmini's ISA (Figure~\ref{fig:synth_res}) in $<$\textbf{1min} and this compiler was implemented in $<$\textbf{100} LOC. Our initial prototype encodes the semantics of matrix multiplication and convolution operator from Gemmini's ISA. Using our initial compiler, we will translate the cloverleaf benchnmarks in \cite{stng} and the image processing kernels introduced in \cite{dexter}.

At present, Metalift can perform code translation for a fixed Gemmini generated accelerator. Our plan is to integrate the search for hardware parameters into the synthesis procedure of Metalift's. The integrated workflow would involve the user providing the source code to be translated and potential Gemmini's generator parameters, such as the scratchpad size or the systolic array dimensions. In addition to searching for equivalent code in Gemmini's ISA, Metalift can also search for the most optimal accelerator parameters for executing the given source program. The search procedure would be guided by the performance of the translated code on the accelerator. However, running every possible candidate on the accelerator may be too expensive and slow down the translation process. Therefore, we need to develop a reliable proxy cost model that can be evaluated quickly and guide Metalift's search procedure. 

A unified framework with Metalift and Gemmini could open up new research directions in improving the hardware software co-design. Overall, this combination offers an exciting opportunity for automatically translating legacy code and achieving high-performance execution on custom hardware.

\bibliographystyle{IEEEtran}
\bibliography{main}

\end{document}